\documentclass[comment, prl,twocolumn,nofootinbib, preprintnumbers, superscriptaddress]{revtex4}

\usepackage{amsmath,amssymb,bm,slashed,braket}
\usepackage{graphicx}
\usepackage{epstopdf}
\usepackage{float}
\usepackage[colorlinks=true,
            linkcolor=blue,
            urlcolor=blue,
            citecolor=green,          
            bookmarks=true,
            bookmarksnumbered=true,
            breaklinks=true,
            pdfpagemode=Fullscreen,
            pdfstartview=FitBH]{hyperref}

\usepackage{tikz} 
\usepackage{tkz-euclide}
\usetikzlibrary{backgrounds} 
\usetikzlibrary{decorations.pathmorphing}
\usetikzlibrary{arrows.meta}
\usetikzlibrary{shapes.misc}
\tikzset{
mystyle/.style={line width=1, baseline, scale=0.6, every node/.style={scale=1}},
v/.style={decorate, draw, decoration={snake, segment length=2.mm, amplitude=0.5mm}},
f/.style={draw, decoration={markings,mark=at position #1 with {\arrow[]{Latex[length=1.5mm,width=1.5mm]}}},
    postaction={decorate},node contents=#1},
f/.default=.6,
fb/.style={draw,decoration={markings,mark=at position #1 with {\arrowreversed[]{Latex[length=1.5mm,width=1.5mm]}}},
    postaction={decorate},node contents=#1},
fb/.default=.6,
s/.style={dashed,draw, decoration={markings,mark=at position #1 with {\arrow[]{Latex[length=1.5mm,width=1.5mm]}}},
    postaction={decorate},node contents=#1},
s/.default=.6,    
sb/.style={dashed,draw,decoration={markings,mark=at position #1 with {\arrowreversed[]{Latex[length=1.5mm,width=1.5mm]}}},
    postaction={decorate},node contents=#1},
sb/.default=.3,
snar/.style={dashed,draw,line width =1.25pt},
cross/.style={cross out, draw=black, minimum size=2*(#1-\pgflinewidth), inner sep=0pt, outer sep=0pt}, 
         }
\usepackage{xcolor}
\definecolor{darkgreen}{rgb}{0,0.5,0}

\usepackage[normalem]{ulem}
\usepackage{color}

\definecolor{Orange}{cmyk}{0,0.61,0.87,0}
\definecolor{JungleGreen}{cmyk}{0.99,0,0.52,0}
\definecolor{OliveGreen}{cmyk}{0.64,0,0.95,0.40}
\definecolor{Brown}{cmyk}{0,0.81,1,0.60}
\definecolor{RoyalBlue}{cmyk}{0.71,0.53,0,0.12}
\definecolor{Gray}{cmyk}{0,0,0,0.40}
\definecolor{LightPink}{cmyk}{0.0,0.25,0,0}
\definecolor{LLightPink}{cmyk}{0.0,0.10,0,0}
\definecolor{LightBlue}{cmyk}{0.25,0,0,0}
\definecolor{LightGray}{cmyk}{0,0,0,0.2}

\usepackage{xcolor}
\definecolor{gesfpurple}{rgb}{0.47,0.19,0.42}

\definecolor{gesflanse}{rgb}{0.00,0.50,0.50}

\definecolor{gesfblue}{rgb}{0.08,0.42,0.76}

\definecolor{gesfred}{rgb}{1,0,0}

\definecolor{gesfwhite}{rgb}{1,1,1}

\definecolor{gesfblack}{rgb}{0,0,0}

\newcommand{\geqn}[1]{Eq.\,\hypersetup{linkcolor=blue}(\ref{#1})\hypersetup{linkcolor=blue}}
\newcommand{\gfig}[1]{{\hypersetup{linkcolor=violet}Fig.\,\ref{#1}\hypersetup{linkcolor=blue}}}

\graphicspath{{figs/}}

\allowdisplaybreaks[4]

\begin{document}

\title{Nucleon Consumption and Mass-Energy Conversion Induced by Dark Matter}

\author{Shao-Feng Ge}
\email{gesf@sjtu.edu.cn}
\affiliation{Tsung-Dao Lee Institute \& School of Physics and Astronomy, Shanghai Jiao Tong University, Shanghai 200240, China}
\affiliation{Key Laboratory for Particle Astrophysics and Cosmology (MOE) \& Shanghai Key Laboratory for Particle Physics and Cosmology, Shanghai Jiao Tong University, Shanghai 200240, China}
\author{Xiao-Dong Ma}
\email{maxid@scnu.edu.cn}
\affiliation{ Key Laboratory of Atomic and Subatomic Structure and Quantum Control (MOE), 
Guangdong Basic Research Center of Excellence for Structure and Fundamental Interactions of Matter, Institute of Quantum Matter, South China Normal University, Guangzhou 510006, China }
\affiliation{Guangdong-Hong Kong Joint Laboratory of Quantum Matter, 
Guangdong Provincial Key Laboratory of Nuclear Science, 
Southern Nuclear Science Computing Center, South China Normal University, Guangzhou 510006, China}

\begin{abstract}
We propose the nucleon consumption induced by dark matter (DM)
as a new scenario to overcome the energy threshold of direct
detection. It can be realized with proton
($\chi + p \rightarrow \chi + \ell^+$) or
neutron ($\chi + n \rightarrow \chi + \nu$) target.
Both effective operators and concrete models are provided
to illustrate the idea. Since the initial DM and nucleon
velocity is only $10^{-3}$ and $1/4$ of the speed of light,
respectively,
the cross section and hence event rate are determined by
the involved particle masses and not affected by the
nucleon Fermi motion. Of the two realizations,
the proton consumption has richer phenomena with both
charged lepton and the daughter nuclei de-excitation
to allow double or even
triple coincidence to significantly suppress the background.
We also illustrate the projected sensitivities at DM and
neutrino experiments such as PandaX-4T, DUNE, JUNO, and
Hyper-K.
\end{abstract}

\maketitle 

{\bf Introduction} --
The nature of dark matter (DM) is one of the biggest unknowns
in our Universe
\cite{Young:2016ala,Arbey:2021gdg,Fairbairn:2022gar}. 
The particle DM is a well-established paradigm that
can explain the observed astrophysical and cosmological
phenomena with minimal parameter set. 
Especially, the weakly interacting massive particle (WIMP), 
whose stability is usually protected by discrete symmetry,
is a theoretically well-motivated scenario that can meet
the required properties to address the DM conundrum 
and at the same time have a detectable possibility
in terrestrial experiments
\cite{Roszkowski:2017nbc,Schumann:2019eaa}.
However, the DM direct detection experiments have not
found the signal but constrain the DM-nucleus
cross section to an unprecedented level for 
the WIMP mass from GeV to hundreds of GeV
\cite{Billard:2021uyg,Misiaszek:2023sxe}.

Since the DM-nucleus elastic scattering can only convert the
DM kinetic energy to the nucleus recoil energy, it is
difficult for a sub-GeV light DM to overcome the energy
threshold of a direct detection experiment.
To solve this issue, novel low-threshold experiments
and theoretical ideas are proposed. 
On the experimental side \cite{Essig:2022dfa,Mitridate:2022tnv}, for instance, 
the DM-electron scattering \cite{Essig:2011nj,Essig:2012yx}, 
the Migdal effect \cite{Ibe:2017yqa}, and 
the bremsstrahlung processes \cite{Kouvaris:2016afs} can
enhance the signal strength. On the theoretical side,
various DM scenarios have appeared.
The detection threshold applies not just for sub-GeV
DM but also for heavier DM whose events
without large enough recoil energy cannot be detected.

The recently proposed fermionic absorption DM is another example of a light DM particle
that can inject enough energy to the detector to produce a detectable signal with either nuclei 
\cite{Dror:2019onn,Dror:2019dib,Li:2022kca,Ma:2024aoc,Ge:2024euk,Ma:2024tkt}
or electron
\cite{Dror:2020czw,Ge:2022ius,Li:2022kca,Ge:2023wye}
target.
The idea is that the DM ($\chi$) is absorbed by nuclear or atomic target ($T$)
while emitting an undetectable neutrino ($\nu$), 
$\chi + T \to \nu + T'$, while the target ($T'$)
recoils. Such signals have been
searched at PandaX-4T \cite{PandaX:2022osq,PandaX:2022ood},
Majorana \cite{Majorana:2022gtu}, EXO-200 \cite{EXO-200:2022adi},
and CDEX \cite{CDEX:2022rxz,CDEX:2024bum}.

Using the fermion absorption and converting its mass
to recoil energy for detecting an unknown particle has
a long history back to the 1940s. 
Pauli also
pointed out the possibility of using its electromagnetic
dipole interactions to probe the neutrino he proposed
in the same paper \cite{Pauli:1991owm}. 
However, both its very existence and electromagnetic
interactions were hypothetical at that time.
With only weak interactions, neutrinos are extremely
difficult to detect. 
So Kan Chang Wang
proposed in 1942
a very smart way of using the $K$-shell electron
capture $e^- + {}^Z_A{\rm N} \rightarrow {}^{Z-1}_{~~A}{\rm N} + \nu$
and detecting the mono-energetic daughter nuclei
recoil to probe the existence of neutrino
\cite{Wang:1942huv,Wang:1947}. This is essentially
absorbing/consuming the known electron and releasing its mass
as energy to probe the unknown neutrino.

In this paper, we propose a new absorption mechanism
by consuming nucleon rather than the DM particle,
to overcome the detection threshold.
Unlike the fermionic absorption DM, which is unstable and suffers from stringent indirect detection constraints due to decaying into SM light particles like neutrinos or photons 
(and possibly electron-positron pair or light mesons were these channels kinematically available)
\cite{Dror:2019onn,Dror:2019dib,Dror:2020czw,Ge:2022ius,Ge:2023wye}, 
the nucleon consuming DM is stable due to a $\mathbb{Z}_2$ symmetry protecting it from decaying. 
Since one nucleon is consumed in the process to release
almost 1\,GeV energy, we anticipate there will be enough energy deposit 
in the detector and thus can be searched for at the DM direct detection
and neutrino experiments.

{\bf Nucleon Consumption} --
If the nucleon consumption happens at the nuclei level,
the target nucleus ($N$) reduces to a lighter one ($N'$),
$\chi + N \rightarrow \chi + N' + \ell/\nu$, where the
final-state charged lepton ($\ell$) or neutrino ($\nu$)
is necessary to balance the
fermion flow and angular momentum. On the other hand, a large
enough momentum transfer would render the nucleon consumption
process to happen at the nucleon level.
Since the DM particle remains intact,
the consumed nucleon in the initial state will become
either a positively charged lepton
($\chi + p \to \chi + \ell^+$) or a neutrino
($\chi + n \to \chi + \nu$).
For both cases, the massive nucleon
(either proton $p$ or neutron $n$)
will release its mass around 938\,MeV to a light charged
lepton (positron $e^+$ or postively charged muon $\mu^+$)
whose mass is much smaller or even a massless neutrino.
With a two-body final state, the momentum transfer
$q^2 \equiv (p_N - p_\ell)^2 = - m_\chi (m^2_N - m^2_\ell) / (m_N + m_\chi)$
is fixed by the DM ($m_\chi$), nucleon ($m_N$), and
lepton ($m_\ell$ for charged lepton or neutrino) masses.
With both $m_\chi$ and $m_N$ around the GeV scale,
the momentum transfer
can probe a length scale much smaller than the nuclei
size. In other words, the consumption scenario
really happens at the nucleon level rather than with
the nucleus as a whole.

For the DM mass below half of the nucleon mass,
the same interaction that induces nucleon consumption
can also lead to the proton
($p \rightarrow 2 \chi + \ell^+$) or neutron
($n\to 2\chi+\nu$) decay. 
To avoid complication, we impose a lower
limit on the DM mass, $m_\chi \gtrsim (m_p, m_n) / 2$,
to guarantee stable nucleons.

With a nonrelativistic incoming DM particle and the
initial proton at rest, the consumed proton mass is
released into the final-state DM and charged lepton
energies $E_\chi = (m^2_p + 2 m_p m_\chi + 2 m^2_\chi - m^2_\ell) / 2 (m_p + m_\chi)$
and $E_\ell = (m^2_p + 2 m_p m_\chi + m^2_\ell) / 2 (m_p + m_\chi)$,
respectively. Both energies are fully determined by the
proton ($m_p$), DM ($m_\chi$), and lepton ($m_\ell$)
masses. Although the
muon mass is not that small, its value is just around
106\,MeV and can hence be safely ignored as a good enough
approximation. The proton
realization can then be detected by the charged lepton
and de-exicitation of the daughter nucleus.

While the charged lepton $e^+$ or $\mu^+$ cannot be
replaced by another charged particle,
the SM neutrino produced by $\chi + n \to \chi + \nu$
will escape detection and can be faked by any other
neutral particles such as the sterile neutrino whose
mass is also small enough or a light dark particle.
Of the two realizations, the neutron consumption with
only the daughter nucleus de-excitation leaves less
signal in the detector. So we would focus on the proton
consumption scenario for illustration while most results
also apply for the neutron case.

The nucleon consumption can be described in terms of
four-fermion effective operators involving a pair of
fermion DM particles, proton, and charged lepton,
\begin{subequations}
\label{eq:O}
\begin{align}
  {\cal O}_{\chi \ell p}^{\tt SS(P)}
& = 
  \overline{\chi}\chi  \, \overline{\ell^c}(i\gamma_5) p, 
\\
  {\cal O}_{\chi \ell p}^{\tt PS(P)}
& = 
  \overline{\chi} i \gamma_5\chi \, 
  \overline{\ell^c} (i\gamma_5) p, 
\\
  {\cal O}_{\chi \ell p}^{\tt VV(A)}
& = 
  \overline{\chi}\gamma_\mu\chi \, 
  \overline{\ell^c} \gamma^\mu (\gamma_5) p, 
\\
  {\cal O}_{\chi \ell p}^{\tt AV(A)}
& = 
  \overline{\chi}\gamma_\mu\gamma_5 \chi \, 
  \overline{\ell^c} \gamma^\mu (\gamma_5) p, 
\\
  {\cal O}_{\chi \ell p}^{\tt TT (\tilde T)}
& = 
  \overline{\chi}\sigma_{\mu\nu}\chi \, 
  \overline{\ell^c} \sigma^{\mu\nu} (i \gamma_5) p, 
\end{align}
\end{subequations}
where $\ell^c$ is the charge conjugation of the lepton field.
For the superscript, the first character stands for the Lorentz
type of the DM current while the second and the one in
parentheses for the proton-lepton current. 
The DM current can also take the fermion number 
violating case,  $\overline{\chi^c}\Gamma \chi$. 
Then only the scalar, pseudo-scalar, and
axial-vector operators
${\cal O}_{\chi \ell p}^{\tt S,P, A}$ can survive. In such a case, additional discrete symmetries can be imposed to forbid unwanted processes such as single nucleon decay, which will be realized in the toy model given below.
The strength of each operator is parametrized by
a Wilson coefficient $C_{\chi \ell p}^i$.  
For the DM-neutron consumption, one may just
replace the charged lepton-proton current by 
the neutrino-neutron current, 
$(\overline{\nu} \Gamma n)$ or $\overline{\nu^c}\Gamma n$.

\begin{figure}[t]
\centering
\includegraphics[width=0.98\linewidth]{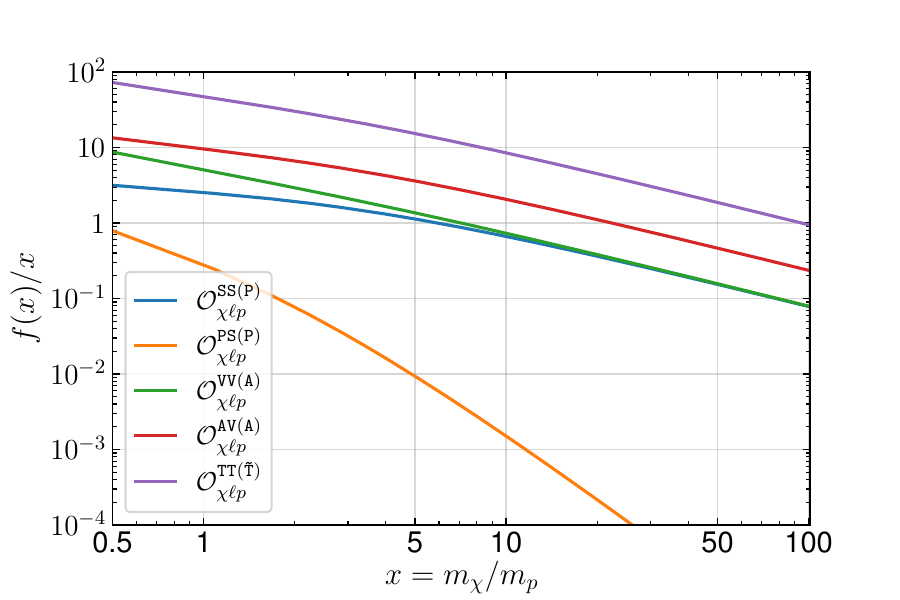}
\caption{The dependence of $f(x)/x$ as a function of $x=m_\chi/m_p$. 
Its value is sizable with the only
exception of the pseudo-scalar case.}
\label{fig:fx}
\end{figure}

\vspace{2mm}
{\bf Consumption at Rest and Fermi Motion} --
Since the halo DM particles travel at only $10^{-3}$
of the speed of light, it is a good approximation
to treat them as at rest. The proton consumption
process then happens at rest which leads to mono-energetic
DM and charged lepton in the final state. Reparametrizing
the DM mass as $x \equiv m_\chi / m_p$, the consumption
cross section takes the form as,
\begin{subequations}
\label{eq:sigmav}
\begin{align}
  \sigma^{\tt SS(P)}_{\chi \ell p} v_\chi
& \approx
  \frac {m^2_p}{32 \pi \Lambda^4_{\tt eff}}
  \frac {(1 + 2 x)^4}
        {(1 + x)^4},
\\
  \sigma^{\tt PS(P)}_{\chi \ell p} v_\chi
& \approx
  \frac {m^2_p}{32 \pi \Lambda^4_{\tt eff}}
  \frac {(1 + 2 x)^2}
        {(1 + x)^4},
\\
  \sigma^{\tt VV(A)}_{\chi \ell p} v_\chi
& \approx
  \frac {m^2_p}{16 \pi \Lambda^4_{\tt eff}}
  \frac {(1 + 2 x)^2 (3 + 4 x + 2 x^2)}
        {(1 + x)^4},
\\
  \sigma^{\tt AV(A)}_{\chi \ell p} v_\chi
& \approx
  \frac {m^2_p}{16 \pi \Lambda^4_{\tt eff}}
  \frac {(1 + 2 x)^2 (3 + 8 x + 6 x^2)}
        {(1 + x)^4},
\\
  \sigma^{\tt TT(\tilde T)}_{\chi \ell p} v_\chi
& \approx
  \frac {m^2_p}{4 \pi \Lambda^4_{\tt eff}}
  \frac {(1 + 2 x)^2 (5 + 10 x + 6 x^2)}
        {(1 + x)^4},
\end{align}
\end{subequations}
where the Wilson coefficients
$C_{\chi e p}^i \equiv \Lambda_{\tt eff}^{-2}$
can be expressed in terms of the effective scale
$\Lambda_{\tt eff}$.
For convenience, one may factorize out the $x$-dependent
part,
$\sigma^i_{\chi \ell p} v_\chi \equiv f^i(x) \times m^2_p / 16 \pi \Lambda^4_{\tt eff}$.
As shown in \gfig{fig:fx}, the $x$-dependent $f(x)/x$
that finally appears in the event rate is typically in the range of $10^{-3} \sim 10$.
The size of $f(x)/x$ is sizable with the
only exception of the pseudo-scalar case.
The typical size of the consumption cross section,
$\sigma_0 \equiv m^2_p / 16 \pi \Lambda^4_{\tt eff}$,
is mainly
determined by the proton mass $m_p$ and the effective scale
$\Lambda_{\tt eff}$.

Since the consumption process happens at the nucleon
level, the Fermi motion of the bound nucleon should be taken
into consideration. The nucleon momentum distribution
in nuclei can be described by the Fermi-gas model
together with a high-momentum tail accounting for the
nucleon-nucleon correlations \cite{Bodek:1980ar}, 
\begin{eqnarray}
  \hat \rho(|\pmb{p}| )
=
  \frac 1 C
\begin{cases}
  1 - 6 \left(k_F a \over \pi \right)^2,
& 0 \leq |\pmb{p}|\leq k_F,
\\ 
  2 R \left(k_F a \over \pi \right)^2 
  \big( {k_F \over |\pmb{p} } |\big)^4,
& k_F < |\pmb{p}|\leq 4\,{\rm GeV},
\\
  0,
& |\pmb{p}|>4\,{\rm GeV},
\end{cases}
\end{eqnarray}
where $a \equiv 2\,{\rm GeV}^{-1}$ is a characteristic size,
$C \equiv {4 \over 3}\pi k_F^3$ the phase space volume,
and $R \equiv 1/( 1- k_F / 4\,{\rm GeV})$, respectively.
The typical Fermi momentum for the xenon nuclei is
$k_F \approx 0.26\, \rm GeV$ which corresponds to
$1/4$ of the speed of light. Nevertheless, the correction
induced by the nucleon velocity $v_N$ to the cross section
only appears at the quadratic order $\mathcal O(v^2_N)$.
Being below 10\,\%, it is safe to omit this correction
for an estimation of the projected sensitivity at DM
and neutrino experiments. Since the dimension-6
four-fermion operators in \geqn{eq:O} scales as
$1/\Lambda^2_{\rm eff}$, the cross section
in \geqn{eq:sigmav} has $1/\Lambda^4_{\rm eff}$
scaling. A 10\,\% deviation in event rate can only
lead to a negligible 2.5\% uncertainty in the projected
sensitivity reach of $\Lambda_{\rm eff}$. Being different
from the fermionic absorption DM case \cite{Ge:2024euk},
the nucleon
consumption cross section is not affected by the nucleon
Fermi motion or Pauli blocking effect since there is
no nucleon in the final state.

\hspace{2mm}
{\bf Search at DM and Neutrino Experiments} --
As mentioned earlier, the proton consumption case will
leave two visible particles, a charged lepton and a daughter
nucleus in the excited state. Since both proton and the DM
particles are heavy with $\mathcal O($GeV) masses, the released
energy is mainly carried away by the light charged lepton.
This also applies for the neutron consumption scenario.
Then the charged leptons with around 500\,MeV
energy can directly deposit energy as
Cherenkov light in a water detector such as Super-K or
Hyper-K as well as scintillation light in others. In
addition, positron $e^+$ can annihilate with a material
electron to create two 511\,keV gamma photons. For
$\mu^+$, it may also decay inside the detector to
produce a Michel positron first. Finally, the
daughter nucleus in the excited state will emit a gamma
photon with typically $\mathcal O$(10\,keV$\sim$MeV)
energy. They can overcome the detection threshold at
DM and neutrino experiments. Altogether, the proton consumption
can leave rich signal phenomena with double or even
triple coincidence that can be used to uniquely select
out the signal and at the same time significantly suppress
backgrounds. It is then safe to assume a background-free
search for a first try of sensitivity estimation.

Probing the proton consumption can be achieved at
DM direct detection experiments such as PandaX, 
XENON, and LZ with low energy thresholds at keV scale.
On the other hand, the neutrino experiments like Super-K,
Hyper-K, and JUNO have much larger size to compensate their
higher thresholds. We take PandaX-4T (4\,ton of ${}^{131}$Xe)
\cite{PandaX:2018wtu}, 
DUNE (40\,kton of ${}^{40}$Ar) \cite{DUNE:2020lwj,DUNE:2020ypp},
JUNO (20\,kton of $\rm C_{18}H_{30}$) \cite{JUNO:2015zny},
and Super-K (22\,kton of H$_2$O) or Hyper-K (190\,kton of H$_2$O)
\cite{Hyper-Kamiokande:2018ofw} to demonstrate the projected sensitivities
at xenon, argon, liquid scintillator, and water based
experiments, respectively. The event rate is,
\begin{eqnarray}
  {d^2N \over d M dt}
= {N_A \times N_{T/p} \over \tilde m_T}
  \rho_\chi
  {\langle \sigma v_\chi \rangle \over m_p x},
\end{eqnarray}
where the DM number density $\rho_\chi / m_\chi$ has been
replaced by $\rho_\chi / m_p x$ which combines with
the $f(x)$ function of $\sigma v_\chi$ to give
$f(x)/x$ as mentioned earlier. Besides the Avogadro number
$N_A=6.022\times 10^{23}$, $N_{T/p}$ is the proton number in
the target $T$ while $\tilde m_T$ is the target molar mass,
\begin{subequations}
\begin{eqnarray}
  N_{\rm {}^{131}Xe/p}= 54,
& \quad
  \tilde m_{\rm {}^{131}Xe} = 
  131.292\, {\rm g/mol},
  \\
  N_{\rm {}^{40}Ar/p}= 18,
& \quad
  \tilde m_{\rm {}^{40}Ar} =39.948 \, {\rm g/mol},
  \\
  N_{\rm C_{18}H_{30}/p}= 138 ,
& \quad
  \tilde m_{\rm C_{18}H_{30}} =246.4308 \, {\rm g/mol}, 
\\
  N_{\rm H_2O/p}= 10,
& \quad
  \tilde m_{\rm H_2O} = 18.02\, {\rm g/mol},
\end{eqnarray}
\end{subequations}
for the four experiments, respectively.
\begin{figure}[t]
\centering
\includegraphics[width=0.98\linewidth]{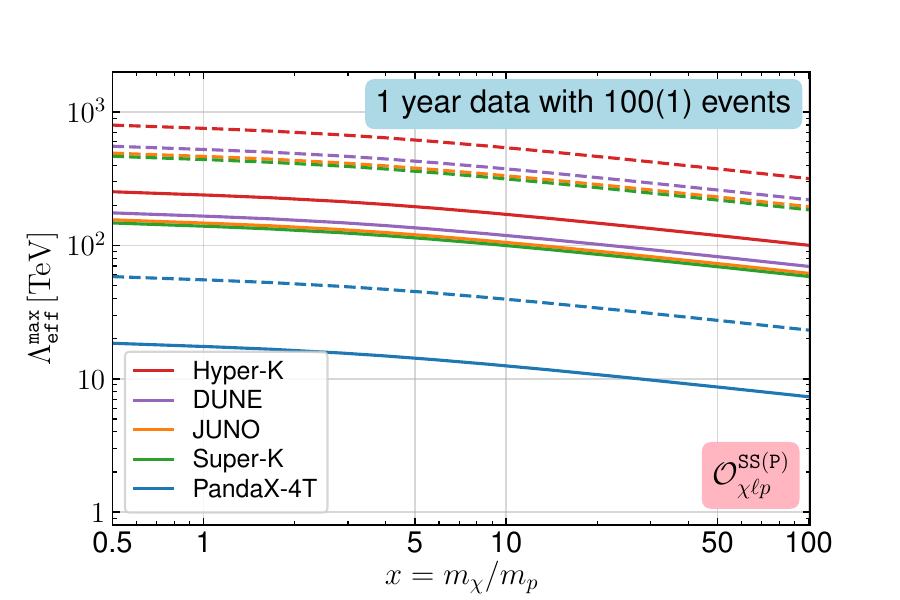}
\includegraphics[width=0.98\linewidth]{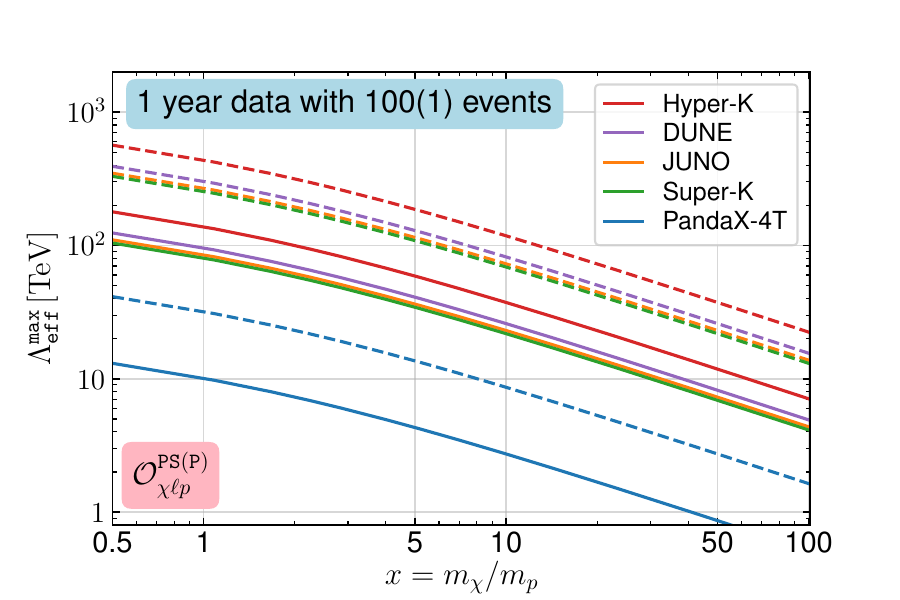}
\caption{The solid (dashed) color curves show the experimental sensitivity on the effective scale $\Lambda_{\tt eff}$ from different experiments by assuming 1 year exposure with 100 (1) events.
The upper (lower) plot is for the interactions with scalar (pseudo-scalar)
DM current.
}
\label{fig:Lambdabound}
\end{figure}
With the DM local density $\rho_\chi = 0.4\, \rm GeV/cm^3$,
the event rates are roughly,
\begin{align}
  {d^2N \over d M dt}
=
\left( 9.2, 7.5, 9.3, 6.8 \right)
  {10^5 \over \rm ton\cdot yr} \left| 1\, \rm TeV  \over \Lambda_{\tt eff}\right|^4
   \frac {f^i(x)} x,
\end{align}
for the different target materials. Considering
the huge volume of 2.25\,kton and 4\,ton for Super-K
and PandaX-4T, the total event number predicted with
$\Lambda_{\tt eff} = 1$\,TeV can reach $10^{10}$
and $10^6$ level within 1 year which is quite significant.
This indicates that the projected sensitivity at DM
and neutrino experiments can be much higher than 1\,TeV.

\gfig{fig:Lambdabound} shows the projected sensitivity
by requiring the signal event number needs to
reach at least 100 (solid) or 1 (dashed). For a light
sub-GeV DM, the effective scale can be probed up to
$\sim 10$\,TeV at the ton-scale PandaX-4T and the
impressive $\gtrsim 200$\,TeV at Hyper-K for 100 events.
In the sub-GeV range, there is no big difference between
the scalar ($\mathcal O^{\tt SS(P)}_{\chi \ell p}$)
and pseudo-scalar ($\mathcal O^{\tt PS(P)}_{\chi \ell p}$)
DM currents since the $x$-dependent function
$f(x)/x$ does not differ much as shown in
\gfig{fig:fx}. With increasing DM mass, the probed
scale of $\Lambda_{\tt eff}$ keeps decreasing since
the $x$-dependent $f(x)/x$ and hence the
event rate keeps decreasing. Between the scalar and
pseudo-scalar DM current operators, the sensitivity
deceases faster for the pseudo-scalar one, 
being consistent with the faster decreases of $f(x)/x$
shown in \gfig{fig:fx}.
The sensitivity for the vector, axial-vector, and tensor
operators is similar to the scalar case with an ${\cal O}(1)$
rescaling factor.

{\bf Toy Model} --
We use the leptoquark model \cite{Dorsner:2016wpm} to provide a concrete
realization of the nucluon consumption operators in \geqn{eq:O}.
Besides a Dirac fermion DM candidate $\chi(1,1,0)$,
the SM is extended with three leptoquarks
$S_1(3,1,2/3)$, $S_2(3,1,-1/3)$,  and $S_3(3,1,-1/3)$, 
where the three numbers in parentheses are
their charges under the SM gauge groups
$SU(3)_c \times SU(2)_L\times U(1)_Y$.
In addition, one needs to impose a discrete
$\mathbb{Z}_4$ symmetry under which the charge
assignments are, 
\begin{align}
 \chi: -i, \quad 
 S_{1,2}: -i,\quad 
 S_3, L, e: -1, 
\end{align}
while the remaining SM particles are $\mathbb Z_4$ singlets.  
Here $L$ and $e$ are the SM left- and right-handed leptons. 
The relevant new physics Lagrangian is 
\begin{align}
{\cal L}_{\tt NP} & \ni  
    y_1 (\overline{u} \chi_R^c)S_1 
  + y_2 (\overline{d} \chi_R^c)S_2
  + y_3 (\overline{Q} \epsilon L^c)S_3
  \nonumber
  \\
  &   + y_4 (\overline{u}e^c)S_3
  - \mu_S \epsilon_{ijk} S_1^i S_2^j S_3^k + {\rm H.c.},
  \label{eq:LNP}
\end{align}
where $Q,u,d$ are the SM quarks while $i,j,k$ are the color
indices in the fundamental representation of $\rm SU(3)_c$. 
For the Yukawa couples $y_{1,2,3,4}$, the flavor indices
have been omitted for simplicity. The dimensional
triple scalar coupling $\mu_S$ is largely a free
parameter.

\begin{figure}[!t]
\centering
\begin{tikzpicture}[mystyle,scale=0.8]
\begin{scope}[shift={(1,-1)}]
\draw[sb, line width =1.4pt, cyan] (-1.73205,1) -- (0,0) node[midway,xshift = 0 pt, yshift = 10 pt]{$S_1$};
\draw[sb, line width =1.4pt, cyan] (1.73205,1) -- (0,0) node[midway,xshift = 0 pt, yshift = 10 pt]{$S_2$};
\draw[sb, line width =1.4pt, orange] (0,-2) -- (0,0) node[midway,xshift = -10 pt, yshift = 0 pt]{$S_3$};
\draw[draw=red,fill=red] (0,0) circle (0.15cm);
\draw[f] (-1.73205,1) -- (-3.09808, 0.633975) node[left]{$u$};
\draw[f,blue] (-1.73205,1) -- (-2.09808,2.36603) node[right]{$\chi$};
\draw[f] (1.73205,1) -- (3.09808, 0.633975) node[right]{$d$};
\draw[f,blue] (1.73205,1) -- (2.09808, 2.36603) node[left]{$\chi$};
\draw[f] (0,-2) -- (-1,-3) node[left]{$Q({\color{magenta!80}u})[{\color{purple}d}]$};
\draw[f] (0,-2) -- (1, -3) node[right]{$L({\color{magenta!80}e})[{\color{purple}\nu_R}]$} ;
\end{scope}
\end{tikzpicture}
\caption{The Feynman diagram for generating the effective operators responsible for the nucleon consumption induced by DM.}
\label{fig:dim9ope}
\end{figure}

The Feynman diagram responsible for the nucleon consuming DM is shown in 
\gfig{fig:dim9ope}, which will generate operators with (pseudo-)scalar DM currents in \geqn{eq:O} 
once quarks condensate into nucleons.
While the couplings with $S_1$ and $S_2$ already contribute
one up and one down quarks, whether the model induces
a proton or neutron consumption is determined by the
$S_3$ Yukawa couplings. A vanishing $y_3$ prevents
neutrino from participating in the process and
hence only leads to the proton consumption scenario
with nonzero $y_4$.
On the other hand, a nonzero $y_3$ can induce both
proton and neutron consumption.
Cancellation between
the $y_3$ and $y_4$ contributions to the proton 
consumption
can also happen by tuning values.
However, it is more desirable
to adjust the model with
a light sterile neutrino $\nu_R(1,1,0)$ that has a
$\mathbb{Z}_4$ charge $-1$. Furthermore,
the SM leptons and quarks all are neutral under the $\mathbb{Z}_4$ symmetry. In this case, the 
Yukawa terms associated with $y_{3,4}$ are replaced by $y_3 (\overline{d}\nu_R^c)S_3$.

The cut-off scale 
$\Lambda_{\tt eff}^{-2}
\sim y_1 y_2 y_{3,4} \mu_S \Lambda_{\tt QCD}^3/ (m_1 m_2 m_3)^2$ is determined by not just the new
physics scale, such as the $S_i$ masses $m_{i}$, but
also the QCD condensation scale $\Lambda_{\tt QCD}$.
The projected sensitivities then convert to a constraint
on the scalar masses,
$m_{i} \lesssim y^{1/2}_{S_i} \mu^{1/6}_S
\Lambda^{1/2}_{\tt QCD} (\Lambda^{\tt max}_{\tt eff})^{1/3}$.
Although the QCD scale is just around $200$\,MeV,
the sensitivity on $m_i$ needs not to be small since
$\mu_S$ can be large.
Currently, the collider bounds on the
leptoquark masses $m_i$ are around 2\,TeV depending
on the decay assumption of leptoquarks
\cite{ATLAS:2020dsk,ATLAS:2024yxs}.
Such constraints translate into a bound on
$\Lambda_{\rm eff}\sim 300\,\rm TeV$ for ${\cal O}(1)$
Yukawa couplings and ${\cal O}(10^{11}\,\rm GeV)$
of $\mu_S$.

{\bf Discussion and Conclusions} --
Not just the DM kinetic energy can induce nucleus
recoil with elastic scattering, the particle mass
can also be absorbed and then converted to the
recoil energy. The previous studies mainly focus on
the dark sector particles. Our study shows that
consuming the nucleons inside a detector target
can provide even more efficient conversion from
mass to the energy deposit. This can overcome not
just the $\mathcal O$(keV) energy threshold
at the DM experiments but also
even the $\mathcal O$(MeV) one at
neutrino experiments. For comparison,
the fermionic absorption DM can only leave a recoil
energy $T_r = m^2_\chi / 2 (m_T + m_\chi)$ where
$m_T$ is the target (electron, nucleon, or nucleus)
mass. Since $m_\chi$ can not be too large ($m_\chi \ll m_N$) such that
the fermionic absorption DM can be stable enough to
survive until today, the recoil energy is actually
much smaller than the DM mass, $T_r \ll m_\chi$.
Since the nucleon consumption
DM mass receives a lower bound, $m_\chi \gtrsim m_p / 2$,
to guarantee the proton stability,
our new scenario of proton consumption can
give the charged lepton an energy of
$E_\ell = (m^2_p + 2 m_p m_\chi + m^2_\ell) / 2 (m_p + m_\chi)$
which needs not to be suppressed and the mass
conversion to deposit energy is very efficient.

\vspace{2mm}
{\bf Notes Added} --
During the finalization of this paper, another paper
\cite{Ema:2024wqr} appeared on arXiv and briefly touched the
similar idea of nucleon consumption DM. Although the
title of their paper emphasized that baryon destruction
is {\it catalyzed} by DM,
the scenario of two-component DM has
a major difference from ours. Our scheme provides
a more economical realization with just a single DM
component. The released mass purely comes
from the consumed nucleon without resorting to the DM
particle.
Although their model can lead to nucleon
consumption with dinucleon initial state, $n+n \to n+ \nu$,
such reaction together with the single nucleon decays like $p \to \pi^0 e^+$ are forbidden in our case due to the exact
$\mathbb{Z}_4$ symmetry. 

The constraint
obtained from the DM capture in neutron star and
the subsequent heating due to nucleon consumption
can also be relaxed. For the neutron
consumption case, the DM capture in neutron star with elastic scattering can be
induced by the mediation of both $S_1$ and $S_2$ where cancellation can happen to make the capture rate small enough.
The neutron star heating effect can be further minimized if the Yukawa terms associated with $y_{3,4}$ in \geqn{eq:LNP} are replaced by $y_3(\bar d \nu_R^c)S_3$ since the right-handed neutrino is sterile and hence cannot deposit energy. 
For the proton consumption case, 
with ${\cal O}(1\sim10)\%$ suppression in the proton number density around the neutron star surface, the heating constraint is less stringent.
In addition, the captured DM needs not to stay around the neutron star surface and can diffuse deep into the neutron star body, which would lead to a further suppression. Since the sizable fraction
of proton density only appears in a very narrow layer around the neutron star surface, such suppression due to DM diffusion is quite significant.

\section*{Acknowledgements}

The authors would like to thank Yong Du for useful discussions
and clarifying the effect of Fermi motion.
SFG is supported by the National Natural Science
Foundation of China (12375101, 12425506, 12090060 and 12090064) and the SJTU Double First
Class start-up fund (WF220442604).
SFG is also an affiliate member of Kavli IPMU, University of Tokyo.
XDM is supported by the Guangdong Major Project of Basic and Applied Basic Research (Grant No.2020B0301030008) and by the NSFC (Grant No. NSFC-12305110).

\bibliography{refs.bib}{}
\bibliographystyle{utphysGe}

\end{document}